\global\def\draftcontrol{0}

   \def\versionno{ crazy modes }
\catcode`\@=11

\expandafter\ifx\csname draftcontrol\endcsname\relax\global\def\draftcontrol{0}
\fi

{\count255=\time\divide\count255 by 60
\xdef\hourmin{\number\count255}
\multiply\count255 by-60\advance\count255 by\time
\xdef\hourmin{\hourmin:\ifnum\count255<10 0\fi\the\count255}}
\def\draftdate{\number\month/\number\day/\number\year\ \ \ \hourmin }

\newcommand\makepapertitle{\par
  \begingroup
    \renewcommand\thefootnote{\@fnsymbol\c@footnote}%
    \def\@makefnmark{\rlap{\@textsuperscript{\normalfont\@thefnmark}}}%
    \long\def\@makefntext##1{\parindent 1em\noindent
            \hb@xt@1.8em{%
                \hss\@textsuperscript{\normalfont\@thefnmark}}##1}%
     \newpage
     \global\@topnum\z@   % Prevents figures from going at top of page.
     \@makepapertitle
     \thispagestyle{empty}\@thanks
  \endgroup
  \setcounter{footnote}{0}%
  \global\let\thanks\relax
  \global\let\makepapertitle\relax
  \global\let\@makepapertitle\relax
  \global\let\@thanks\@empty
  \global\let\@author\@empty
  \global\let\@date\@empty
  \global\let\@title\@empty
  \global\let\title\relax
  \global\let\author\relax
  \global\let\date\relax
  \global\let\and\relax
  \def\version{\let\version\@version\@gobble}
}
\def\@makepapertitle{%
  \newpage
   \ifnum\draftcontrol=1 {}
   \version\versionno
   \vskip 3em%
   \else
   \hfill\hbox to 3cm {\parbox{4cm}{\@pubnum}\hss}%
   \vskip 3em%
   \fi
   \begin{center}%
   \let \footnote \thanks
     {\LARGE {\@title}}%
     \vskip 1.5em%
     {\normalsize%\large
       \lineskip .5em%
       \begin{tabular}[t]{c}%
         \@author
       \end{tabular}\par}%
     \vskip 1.5em%
     {\@bstract}%
     \end{center}%
     \vskip 1.5em
     \@date%
   \par
}

\gdef\@pubnum{}
\def\pubnum#1{%
  \gdef\@pubnum{#1}}

\gdef\@bstract{}
\def\Abstract#1{%
  \gdef\@bstract{%
   \parbox{\textwidth-0pc}{%
   \centerline{\bf Abstract}\penalty1000%
\kern.2cm%
\noindent%\abstractfont \baselineskip=12pt
\renewcommand\baselinestretch{1.0}%
{#1}}}
}

\def\ps@paper{\let\@mkboth\@gobbletwo%
     \ifnum\draftcontrol=1
    \def\@oddfoot{\hbox to \textwidth{\tiny \versionno \hfil\tiny\draftdate}%
    \hskip -\textwidth \hbox to \textwidth{\hfil\rm\thepage\hfil}}%
     \else\def\@oddfoot{\hbox to \textwidth{\hfil\rm\thepage\hfil}}
     \fi
     \let\@evenfoot\@oddfoot
}
\def\body{\clearpage
          \pagestyle{paper}
    }
\def\@version#1{\ifnum\draftcontrol=1
\typeout{}\typeout{#1}\typeout{}
\vskip3mm\centerline{\hbox{\fbox{\normalsize{\tt DRAFT -- #1 -- }
                   {\draftdate}}}}\vskip3mm
\fi}
\let\version\@version
\long\def\eqlabel#1{\ifnum\draftcontrol=1
                    \tag@false  % there are some problems with multline without this
                    \tag*{(\theequation) \hbox to -0.2cm{\hspace{0cm}\small{#1}\hss}}
                    \refstepcounter{equation}
                    \edef\@currentlabel{\theequation}
                    \ltx@label{#1}          % use old LaTeX \label instead of new definition
                    \else
                    \label{#1}
                    \fi
                    }
\let\st@bibitem\@bibitem
\let\st@lbibitem\@lbibitem
\ifnum\draftcontrol=1
  \def\@bibitem#1{%
    \st@bibitem{#1}\a@@label{#1}\ignorespaces}
  \def\@lbibitem[#1]#2{%
    \st@lbibitem[#1]{#2}\a@@label{#2}\ignorespaces}
  \def\a@@label#1{%
    \gdef\a@lab{\smash{\normalfont\small#1}}
    \ifvmode
      \if@inlabel
        \global\setbox\@labels\hbox{%
          \llap{\a@lab\let\a@lab\relax
                \kern\@totalleftmargin\kern\marginparsep}%
          \box\@labels}%
      \fi
    \fi}
\fi
\documentclass[12pt,letterpaper]{article}
\usepackage{amsmath,amssymb,array,calc,epsfig}
\ifnum\draftcontrol=1
\fi
\renewcommand\baselinestretch{1.25}
\renewcommand\section{\@startsection {section}{1}{\z@}%
                                   {-3.5ex \@plus -1ex \@minus -.2ex}%
                                   {2.3ex \@plus.2ex}%
                                   {\normalfont\large\bfseries}}
\renewcommand\subsection{\@startsection{subsection}{2}{\z@}%
                                   {-3.25ex\@plus -1ex \@minus -.2ex}%
                                   {1.5ex \@plus .2ex}%
                                   {\normalfont\normalsize\bfseries}}
\renewcommand\subsubsection{\@startsection{subsubsection}{3}{\z@}%
                                   {-3.25ex\@plus -1ex \@minus -.2ex}%
                                   {1.5ex \@plus .2ex}%
                                   {\normalfont\normalsize\it}}
\renewcommand\paragraph{\@startsection{paragraph}{4}{\z@}%
                                   {-3.25ex\@plus -1ex \@minus -.2ex}%
                                   {1.5ex \@plus .2ex}%
                                   {\normalfont\normalsize\bf}}

\def\ie{{\it i.e.}}

\def\revise#1       {\raisebox{-0em}{\rule{3pt}{1em}}%
                     \marginpar{\raisebox{.5em}{\vrule width3pt\
                     \vrule width0pt height 0pt depth0.5em
                     \hbox to 0cm{\hspace{0cm}{%
                     \parbox[t]{4em}{\raggedright\footnotesize{#1}}}\hss}}}}

\def\calo         {{\cal O}}

\def\del          {\partial}

\def\sqr#1#2{{\vcenter{\vbox{\hrule height.#2pt
 \hbox{\vrule width.#2pt height#1pt \kern#1pt
 \vrule width.#2pt}\hrule height.#2pt}}}}

\def\a{\alpha}

\newcommand{\qq}{\mathfrak{q}}
\newcommand{\ww}{\mathfrak{w}}

\def\e{\epsilon}
\def\r{\rho}

\def\w{\omega}

\newcommand{\be}{\begin{equation}}

\newcommand{\ee}{\end{equation}}
\newcommand{\bea}{\begin{eqnarray}}
\newcommand{\eea}{\end{eqnarray}}

\def\hs{\hat{\sigma}}
\def\ts{\tilde{\sigma}}

\catcode`\@=12

\begin{document}

%%%
%%%%%% text starts here
%%%%%%%%%

\title{\bf Relativistic Conformal Magneto-Hydrodynamics from Holography}
\pubnum{UWO-TH-09/4}

%\date{November 2008}

\author{
Evgeny I. Buchbinder$ ^{1}$,  Alex Buchel$ ^{1,2}$ \\[0.4cm]
\it $ ^1$Perimeter Institute for Theoretical Physics\\
\it Waterloo, Ontario N2J 2W9, Canada\\
\it $ ^2$Department of Applied Mathematics\\
\it University of Western Ontario\\
\it London, Ontario N6A 5B7, Canada
}

\Abstract{
We use the AdS/CFT correspondence to study first-order relativistic viscous magneto-hydrodynamics 
of (2+1) dimensional conformal magnetic fluids. It is shown that the first order 
magneto-hydrodynamics constructed following Landau and Lifshitz from the 
positivity of the entropy production 
is inconsistent. We propose additional contributions to the entropy motivated dissipative current
and, correspondingly, new dissipative transport coefficients. We use the strongly coupled M2-brane 
plasma in external magnetic 
field to show that the new magneto-hydrodynamics leads to self-consistent results in the shear and  
sound wave channels.
}

\makepapertitle

\body

\version\versionno

%%%%%%%%%%%%%%%%%%%%%%%%%%%%%%%%%%%%%%%%%%%%%%%%%%%%%%%%%%%%%%%%%%%%%%%%%%%%%%%%%%%%%%%%%%%%%%%%%%%%%%
A theory of viscous magneto-hydrodynamics \cite{LL} 
has a wide range of practical applications in condensed matter \cite{Kovtun} (and references therein)
and astrophysics \cite{astro} (and references therein). It is best thought as an effective 
low-energy theory organizing  derivative expansion
of systems in local, but not global, equilibrium. For concreteness, 
consider relativistic $d$ dimensional charged fluid in 
external electro-magnetic 
field $F^{\mu\nu}$. 
The conserved quantities associated with the system are the stress-energy 
tensor $T^{\mu\nu}$ and the electric current $d$-vector
$J^\mu$. The dynamics of the hydrodynamic fluctuations in the fluid 
is governed by the following equations of motion
\begin{equation}
\del_\mu J^\mu=0\,,\qquad \del_\nu T^{\mu\nu}=F^{\mu\nu}J_\nu\,, 
\label{eoms}
\end{equation}   
where the total current and the stress-energy tensor include the equilibrium 
part (with local energy density $\epsilon$, pressure $P$,
and charge density $\rho$)
and the dissipative 
parts ($\nu_\mu$ and $\Pi^{\mu\nu}$ correspondingly):
\begin{equation}
T^{\mu \nu} =\epsilon u^{\mu}u^{\nu}+P \Delta^{\mu \nu} +\Pi^{\mu \nu}\,, 
\qquad J^{\mu}=\rho u^{\mu} +\nu^{\mu}\,.
\label{2.11}
\end{equation}
In Eq.~\eqref{2.11} 
\begin{equation}
\begin{split}
\Delta^{\mu \nu}=\eta^{\mu \nu}+u^{\mu}u^{\nu}\,,\;\;\; \Pi^{\mu}_{\;\;\nu} u^\nu = 0\,,\;\;\; u^{\mu}\nu_{\mu}=0\,,
\end{split}
\label{somedef}
\end{equation} 
where $u^\mu$ is the local $d$-velocity of the fluid, $u^\mu u_\mu=-1$. 
In phenomenological hydrodynamics one represents the 
dissipative terms $\nu_\mu$ and $\Pi^{\mu\nu}$ as an infinite series expansion in 
velocity gradients, with coefficients of 
the expansion commonly referred to as transport coefficients. 
A familiar example are the (neutral) fluid Navier-Stokes equations, obtained 
by truncating the dissipative stress-energy tensor
$\Pi^{\mu\nu}$ to linear order in the velocity gradients
\begin{equation}
\Pi^{\mu \nu}=-\eta \sigma^{\mu \nu} -\zeta \Delta^{\mu \nu} (\partial_{\alpha} u^{\alpha})\,, 
\label{2.14}
\end{equation}
where 
\begin{equation}
\sigma^{\mu \nu}=\Delta^{\mu \alpha} \Delta^{\nu \beta}
(\partial_{\alpha}u_{\beta}+\partial_{\beta}u_{\alpha})
-\frac{2}{d-1}\Delta^{\mu \nu}\Delta^{\alpha \beta}(\partial_{\gamma}u^{\gamma})\,.
\label{2.15}
\end{equation}
Notice that at this order in the hydrodynamic approximation we need to 
introduce only two transport coefficients: 
the shear $\eta$ and bulk $\zeta$ viscosities. Higher orders in the velocity 
gradients expansion for $\Pi^{\mu\nu}$, and for charged  
fluids the dissipative current velocity gradients expansion, 
necessitate introduction of additional transport coefficients.
It is desirable to formulate a principle that would constrain such a rapidly 
growing set of transport coefficients.   
A popular approach, due to Landau and Lifshitz \cite{LL}, is to define an 
off-equilibrium entropy current and constrain 
transport coefficients by imposing the positivity of its 
divergence (equivalently, the positivity of the entropy 
production in relaxation of a near-equilibrium state). In this framework the 
authors of~\cite{Kovtun} (HKMS)
determined the first-order in $d$-velocity gradients expression for the dissipative current $\nu_\mu$:
\begin{equation}
\nu^{\mu}=\sigma_{Q}\Delta^{\mu \nu}(-\partial_{\nu}\mu
+F_{\nu \alpha}u^{\alpha}+\frac{\mu}{T}\partial_{\nu}T)\,,
\label{2.17}
\end{equation}
where $T$ and $\mu$ and the local temperature and the chemical potential.
Remarkably, only a single new transport coefficient (in addition to $\{\eta,\zeta\}$), 
a conductivity  $\sigma_Q$, is needed to describe first-order HKMS 
magneto-hydrodynamics.
It is important to stress that in~\cite{Kovtun}, 
the authors considered the case of the electro-magnetic field 
vanishingly small in the hydrodynamic limit, that is 
$F \sim {\cal O}(\partial u)$. 

The question that we raise in this Letter is whether 
the principle of the positivity of the entropy production
leads to a consistent theory of magneto-hydrodynamics
for an arbitrary external electro-magnetic field, not necessarily 
small in the hydrodynamic limit.  
Unless there is a first principle microscopic
derivation of hydrodynamics, it is impossible to address such a question.  
In this Letter we consider the three-dimensional 
maximally supersymmetric 
Yang-Mills theory with large number of colors
at the infrared fixed point. This theory is commonly referred to as the theory of 
M2-branes. 
Then we use 
AdS/CFT correspondence~\cite{Juan} of the strongly coupled M2-brane plasma 
in external magnetic field to argue that  
the first-order dissipative current $\nu_\mu$ has to be modified as follows
\begin{equation}
\begin{split}
 \nu^{\mu}&=\sigma_{Q}\Delta^{\mu \nu}(-\partial_{\nu}\mu
+F_{\nu \alpha}u^{\alpha}+\frac{\mu}{T}\partial_{\nu}T) \\
&- {\hs_Q} \Delta^{\mu\nu} F_{\nu\a} u^\gamma \del_\gamma (\frac {u^\a}{T})\,,
\end{split}
\label{new}
\end{equation}
where $\hs_Q$ is the new conductivity transport 
coefficient. Given that AdS/CFT correspondence provides a microscopic derivation of the 
M2-brane hydrodynamics, we compute that in this particular case 
and at zero charge density at equilibrium 
\begin{equation}
\hs_Q=\frac{3}{2\pi}\ \sigma_Q\,.
\end{equation}
This implies that the principle of the positivity of the entropy 
production is violated in magneto-hydrodynamics already at {\it the linearized level}
as the theory of M2-branes represents an explicit counter-example.
On the contrary, in neutral hydrodynamics with no external fields this 
principle holds to linear order and is violated at second order~\cite{Baier}.

Our approach to introducing the transport coefficients to describe the
dissipative current $\nu^{\mu}$ in conformal magneto-hydrodynamics
will be based on conformal invariance.
A similar approach to conformal hydrodynamics
with no extra charges and fields was undertaken in~\cite{Baier}.
Under the Weyl rescaling with a local 
parameter $\Omega$
we have the following 
transformation properties 
\begin{equation}
\begin{split}
& 
T^{\mu \nu}\to e^{(d+2)\Omega}T^{\mu \nu}\,, \quad
J^{\mu}\to e^{d \Omega}J^{\mu}\,, \quad
F_{\mu \nu} \to F_{\mu \nu}\,,
\\
& \epsilon \to e^{d \Omega} \epsilon\,, \quad T\to e^{\Omega}T\,, \quad u^{\mu} \to e^{\Omega} 
u^{\mu}\,,\quad
\mu \to  e^{\Omega} \mu\,.
\end{split}
\label{e1}
\end{equation}
These properties are obtained from demanding that eqs.~\eqref{eoms}
are Weyl invariant. Based on the principle of conformal invariance as well as 
transversality $u_{\mu} \nu^{\mu}=0$, 
we propose the following expression for the 
dissipative part of the current 
up to terms which do 
not contribute to linear order in fluctuations. We get
\begin{equation}
\begin{split}
\nu^{\mu}=&\sigma_{Q 1}\Delta^{\mu \nu}(\frac{\mu}{T}\partial_{\nu}T-\partial_{\nu}\mu)
+\sigma_{Q 2}\Delta^{\mu \nu}F_{\nu \alpha} u^{\alpha} \\
&-
\hat{\sigma}_Q\Delta^{\mu \nu}F_{\nu \alpha}u^{\gamma}\partial_{\gamma}(\frac{u^{\alpha}}{T})\,.
\end{split}
\label{e2}
\end{equation}
We have introduced the two different coefficients $\sigma_{Q 1}$ and $\sigma_{Q 2}$ 
instead of $\sigma_{Q}$ since the first two terms individually are conformally invariant. 

Let us discuss whether there are possible corrections to~\eqref{e2}.
Our discussion will be rather brief since these extra terms will not 
play any role in the rest of the Letter. 
First, 
there is one more term linear in $F$ which we can add in~\eqref{e2}, namely
\begin{equation}
\ts_Q\Delta^{\mu \nu}F_{\nu \alpha} (\frac{\mu}{T}\partial^{\alpha}T-\partial^{\alpha}\mu)\,.
\label{e3}
\end{equation}
In the rest of the paper, we will consider the neutral
M2-brane plasma and set $\rho=0$. 
In this case, as we will see below, there is a certain decoupling
of the linearized hydrodynamic equations.
The term~\eqref{e3} is inconsistent with the decoupling and, hence, 
$\ts_Q=0$ in this particular case. Note that, in general, $\ts_Q=0$ 
does not have to vanish and the term~\eqref{e3} 
can appear in~\eqref{e2}. However, it vanishes 
for the theory under study. Similarly, one can analyze terms with higher
powers of $F_{\mu \nu}$. It is not difficult to show that such terms either are 
inconsistent with the decoupling (so the coefficient in front of them 
has to vanish just like $\ts_Q=0$) or can be used to redefine the transport 
coefficients in~\eqref{e2} at higher order of $F^2$. The latter is due 
to the fact that in pure magnetic background in three dimensions 
$F_{\mu}^{\alpha} F_{\alpha \nu}\sim F^2 \eta_{\mu \nu}$. 
Thus, we would like to stress that our proposal in eq.~\eqref{e3} 
is {\it not} the most general expansion of the current in the external 
electro-magnetic field. As we just explained there can be additional terms
invisible by the magnetized M2-brane theory plasma.
The purpose of this Letter is not to develop 
such an expansion but rather to explicitly demonstrate that the principle 
of the positivity of the entropy production in violated in magneto-hydrodynamics
at the linearized level.

We would like now to test our phenomenological proposal of first order magneto-hydrodynamics 
in the framework of soluble
AdS/CFT correspondence \cite{Juan}. Specifically, 
we consider magneto-hydrodynamics of dyonic black holes in $AdS_4\times S^7$
supergravity of M-theory. The latter realizes a holographic dual of the
strongly coupled conformal $(2+1)$ dimensional magnetized plasma. 
The equilibrium state of the plasma is described by a dyonic black hole in the 
effective four-dimensional Einstein-Maxwell gravity with a negative 
cosmological constant \cite{HK}. A thermodynamic potential of the plasma is given by 
\begin{equation}
\Omega =-V_2\ p= V_2\ \frac{1}{g^2}   
\frac{\a^3}{4} \left(-1-\frac{\mu^2}{\a^2}+3\frac{B^2}{\a^4}\right)\,,
\label{th1}
\end{equation}
where $V_2$ is the spatial area and
$p$ is the thermodynamic pressure. Furthermore, 
$g^2$ is the bulk gravitational coupling, 
related to the central charge $c$ of the M2-brane CFT as 
$c=96 g^{-2}$ (it can be computed from eq. (5) in~\cite{Ritz}),
$\mu$ is the electric charge chemical potential, 
and $B$ is an external magnetic field. The parameter $\a$ 
entering \eqref{th1} is related to the temperature $T$ as follows
\begin{equation}
\frac{4\pi T}{\a}=3-\frac{\mu^2}{\a^2}-\frac{B^2}{\a^4}\,,
\label{hqa}
\end{equation} 
and is introduced to simplify expressions for further thermodynamic quantities. 
For example, the equilibrium energy density $\e$,
the entropy density $s$, the charge density $\r$ and the magnetization $M$ per unit areas are given by
\begin{equation}
\begin{split}
&\e=\frac{1}{g^2} \ \frac{\a^3}{2}
\left(1+\frac{\mu^2}{\a^2}+\frac{B^2}{\a^4}\right)\,,\quad s=\frac{1}{g^2}  \a^2\,,\\
&\r=\frac{1}{g^2}  \a\mu\,,\quad 
M=
-\frac{1}{g^2}  \frac{B}{\a}\,.
\end{split}
\label{th2}
\end{equation}
Furthermore, the external magnetic field $B$ naturally enters 
the supergravity analysis via the parameter $h=\frac {B}{\a^2}$.

We denote $(x^0=t, x^1=x, x^2=y)$ and refer to nonzero 
components of the electromagnetic field strength as $F_{x y}=-F_{y x}=B$. 
Eq.~\eqref{e2} for the dissipative current 
together with eqs.~\eqref{eoms}-\eqref{2.15} is our starting point.
Before we discuss the linearized equations of motion, let us point out 
that the appearance of the magnetic field substantially increases the possible set of the hydrodynamic
regimes which one can study. That is, instead of  taking $B$ to be fixed
in the hydrodynamic limit of small momentum $q$, we can simultaneously scale $B$ as  $q^p$, $p>0$.
As explained in \cite{bbv,bb}, the new hydrodynamic regimes allow us to study effects of magnetic field on the 
dispersion relation of the sound  and the shear quasinormal modes in a controllable setting, \ie, 
when it is consistent to neglect the second and higher order hydrodynamic 
contributions, as well as nonlinear effects.  
In this Letter, we will study the cases of fixed $B$ and $B \sim q^{1/2}$. This will allow us to extract
the coefficients $\sigma_{Q 1}$, $\sigma_{Q 2}$ and $\hat{\sigma}_Q$ for the M2-brane plasma.

We expand eqs.~\eqref{eoms} around the equilibrium state 
$u^{\mu}=(1, 0, 0)$, $T=const$, $\mu=0$ to linear order in fluctuations. As usual, the 
fluctuations are of the plane wave form $e^{-i \omega t + i q y}$. As the set of independent 
fluctuations we can choose $(\delta \epsilon, \delta \mu, \delta u_x, \delta u_y)$. 
In deriving the equations of motion, in addition to setting $\rho=0, \mu=0$, we will also set
(see eqs.~\eqref{th2})
\begin{equation}
\left(\frac{\partial \rho}{\partial T}\right)_{\mu}=0\,, \qquad 
\left(\frac{\partial \epsilon}{\partial \mu}\right)_{T}=0\,, \qquad 
\left(\frac{\partial \rho}{\partial \mu}\right)_{T}=\frac{1}{g^2} \a\,. 
\label{e6}
\end{equation}
At $\rho=0$, the equations of motion decouple into the two separate pairs. 
The first pair reads
\begin{equation}
\begin{split}
& 0=\w \delta\epsilon - \frac{3}{2} q \e \delta u_y\,, \\
&0= \frac{3}{2}
\w \e \delta u_y -\frac{1}{2}q \delta \e
+i q^2 \eta \delta u_y+ i \sigma_{Q2} B^2 \delta u_y
\\&-\frac{\w \hat{\sigma}_Q B^2 \delta u_y}{T}
\,,
\end{split}
\label{e7}
\end{equation}
where we have used the equation of state $P=\e/2$ followed from conformal invariance.
When $B=0$ these two equations describe sound waves and we will refer to them as to the 
``sound channel''.
The second pair of equations is
\begin{equation}
\begin{split}
&0=\frac{3}{2} \w \e \delta u_x -q B \sigma_{Q1} \delta \mu +
i \sigma_{Q2} B^2 \delta u_x +i q^2 \eta \delta u_x
\\
&-\frac{\w \hat{\sigma}_Q B^2 \delta u_x}{T}
\,,\\
&0= \w \left( \frac{\del \rho}{\del \mu}\right)_{T} \delta \mu +
q \sigma_{Q2} B \delta u_x +i q^2 \sigma_{Q1} \delta \mu
\\
&+\frac{ i \w q \hat{\sigma}_Q B \delta u_x}{T}
\,.
\end{split}
\label{e8}
\end{equation}
When $B=0$ these two equations further decouple. One of them describes a shear 
mode and the other one describes charge diffusion. We will refer to this pair
as to the ``shear channel''. The decoupling between the sound and the shear channels
was also observed in the supergravity description of the M2-brane plasma with 
vanishing equilibrium charge density~\cite{bbv}. 
Now note that the term~\eqref{e3} would lead to the mixing between the sound 
and shear channels. Hence, $\ts_Q=0$ at least at $\rho=0$.
Unfortunately, at this point, 
we are not able to say whether the term~\eqref{e3} is present in general in magneto-hydrodynamics. 

Let us now discuss solutions to eqs.~\eqref{e7} and~\eqref{e8} in the regimes specified above. 
Recall that 
in the context of AdS/CFT correspondence hydrodynamic fluctuations in plasma,
\ie, the sound modes and the shear modes, are related to the lowest quasinormal modes of the black hole solutions dual to the 
equilibrium state of the plasma \cite{ss}. For the magnetized M2-brane plasma of interest here, these quasinormal modes 
were extensively studied in \cite{bbv,bb}. For convenience, we introduce $\ww=\omega/(2\pi T)$ 
and $\qq=|\vec q|/(2\pi T)$.
\\
{\it Hydrodynamic limit with $B$ held fixed}.
In the sound channel we obtain a constant cyclotron mode and a diffusive mode 
\begin{equation}
\w^{sound}_{CFT} =-i q^2 \frac{3 \e}{4 \sigma_{Q 2} B^2}\,.
\label{e9}
\end{equation}
This mode was reproduced on the supergravity side in~\cite{bb}
where the comparison has also been made. The result of this comparison is that 
\begin{equation}
\sigma_{Q 2}=\frac{1}{g^2}\,.
\label{e10}
\end{equation}
It is important to stress that this result is exact to {\it all orders} in magnetic field. 
In the shear channel 
we obtain a cyclotron mode and a subdiffusive mode 
\begin{equation}
\w^{shear}_{CFT} =-i q^4 \frac{\eta}{B^2 
\left(\frac{\partial \rho}{\partial \mu}\right)_T} 
\frac{\sigma_{Q 1}}{\sigma_{Q2}}\,.
\label{e19}
\end{equation}
This mode was reproduced on the supergravity side in~\cite{bb}. Moreover, in~\cite{bb}
$\eta$ was found from a Kubo formula to all orders in magnetic field
to satisfy
\begin{equation}
\frac{\eta}{s}=\frac{1}{4 \pi}\,.
\label{e20}
\end{equation}
It was further generalized for the case of $\rho\neq 0$ in~\cite{Hansen}.
Comparison with the supergravity result of~\cite{bb} then implies that 
$\sigma_{Q 1} =\sigma_{Q2}$. From now on we will remove the subscripts ``1'' and ``2''
and denote $\sigma_{Q 1} =\sigma_{Q2}=\sigma_Q$. Strictly speaking we find that
$\sigma_{Q 1} =\sigma_{Q2}$ only for the M2-brane plasma. However, it is 
possible that this is true in general.
\\
{\it Hydrodynamic limit with $H\equiv \frac{h}{\sqrt{\qq}}$ held fixed}.
The relevant dispersion relation for the sound channel quasinormal  mode of dyonic black hole describing strongly coupled 
M2-brane plasma was computed in \cite{bbv}:
 \begin{equation}
\begin{split}
&\ww^{sound}_{SUGRA}=\Gamma_0^{sound}\ \qq+ i\Gamma_1^{sound}\ \qq^2+\calo\left(\qq^3\right)\,,\\
&\Gamma_0^{sound}=-i\ \frac 49 H^2\pm \frac {1}{\sqrt 2}\sqrt{1-\frac{32}{81}H^4}\,,\\
&\Gamma_1^{sound}=-\frac 14-\frac 89H^4\pm\frac{2iH^2(32 H^4-45)}{9\sqrt{162-64 H^4}}\,.
\end{split}
\label{sound}
\end{equation}

On the CFT side, the dispersion relation in the sound channel now depends on the new coefficient 
$\hs_Q$:
\begin{equation}
\begin{split}
&\ww^{sound}_{CFT}=\Gamma_0^{sound}\ \qq+ i\Gamma_1^{sound}\ \qq^2\\
&+H^2 \left(\frac {32}{27}H^2\mp\frac{4i(64 H^4-81)}{27\sqrt{162-64 H^4}}\right)\left(1-\frac{2\pi}{3}\frac{\hs_Q}{\sigma_Q}\right)\qq^2\\
&+\calo\left(\qq^3\right)\,,
\end{split}
\label{cftsound}
\end{equation}
where we used equilibrium thermodynamic properties of the M2-brane plasma, eqs.~\eqref{th2},
the universality of its viscosity, eq.~\eqref{e20}, and eq.~\eqref{e10}. 
If we assume that the principle of the positivity of the entropy production 
holds to linear order and set $\hs_Q$ to zero we find that 
\begin{equation}
\ww^{sound}_{SUGRA}-\ww^{sound}_{CFT} =\calo(\qq^2)\ne \calo(\qq^3)\,.
\label{disc}
\end{equation}
This shows that the formulation of 
magneto-hydrodynamics based on this principle is {\it inconsistent}.
Moreover, the inconsistency appears already at the linearized level.
In our framework of new dissipative 
magneto-hydrodynamics \eqref{e2} the inconsistency is resolved by simply 
declaring that 
%\cite{com1}
%
\begin{equation}
\hs_Q=\frac{3}{2\pi}\ \sigma_Q \,.
\label{hs}
\end{equation}
Note that, formally, this relation is computed only to leading order in
$B^2/T^4$ since our analysis is done in the hydrodynamic limit 
with the vanishing magnetic field. However, given that $\sigma_Q$ is independent of $B^2/T^4$,
it is natural to expect that so is $\hs_Q$. 

Can we find an independent check on relation \eqref{hs} 
between the two conductivities of the M2-brane plasma? 
Fortunately, the answer is {\it yes}! 
Consider the shear channel hydrodynamic regime with $H=\frac{h}{\sqrt{\qq}}$ held fixed. 
On the CFT side the dispersion relation again depends on the new coefficient $\hs_Q$:
\begin{equation}
\begin{split}
\ww^{shear}_{CFT}=&-i \frac 89 H^2\ \qq+i
\left(-2+\left(\frac{16}{27}-\frac{128\pi}{81}\frac{\hs_Q}{\sigma_Q}\right)H^4\right)\qq^2\\
&+\calo(\qq^3)\,, 
\end{split}
\label{cftshear}
\end{equation}
where once again, we used equilibrium thermodynamic properties of the M2-brane plasma, eqs.~\eqref{th2},
the universality of its viscosity, eq.~\eqref{e20}, and eq.~\eqref{e10}. 

Although the shear channel analysis in \cite{bb} were not done in the  
hydrodynamic regime with $B\sim q^{1/2}$, it is straightforward to 
do so. Following  notations of \cite{bb}, the gauge-invariant wave-functions 
of the shear channel quasinormal modes $\{Z_H, Z_A\}$,
in the hydrodynamic limit with  $\frac{h}{\sqrt \qq}$ kept constant, 
have the following hydrodynamic expansion
\begin{equation}
\begin{split}
Z_H (r)=& f(r)^{-i \ww/2} ( Z_0(r)+ i \qq Z_1(r)+ \qq^2 Z_2(r)
+ 
{\cal O}(\qq^3) )\,,
\\
Z_A (r)=& f(r)^{-i \ww/2} \a H\sqrt{\qq}  ( A_0(r)+ i \qq A_1(r)+ \qq^2 A_2(r)\\
&+ {\cal O}(\qq^3))\,.
\end{split}
%\label{3.9}
\end{equation}
As in \cite{bb}, imposing the incoming wave boundary conditions 
at the horizon and the Dirichlet conditions at the boundary for $\{Z_H,Z_A\}$
determines the dispersion relation for the hydrodynamic shear mode  
\begin{equation}
\begin{split}
&\ww^{shear}_{SUGRA}=\Gamma_0^{shear}\ \qq+ i\Gamma_1^{shear}\ \qq^2+\calo\left(\qq^3\right)\,,\\
&\Gamma_0^{shear}=-i\ \frac 89 H^2 \,,\qquad 
\Gamma_1^{shear}=-2-\frac{16}{9} H^4\,.
\end{split}
\label{shear}
\end{equation}
Requiring that 
\begin{equation}
\ww^{shear}_{SUGRA}-\ww^{shear}_{CFT}= \calo(\qq^3)
\label{shear1}
\end{equation}
\\
we arrive at the same value for $\hs_Q$, as the one obtained in the sound channel, eq.~\eqref{hs}.

In this Letter, we presented a new phenomenological framework of 
first-order viscous magneto-hydrodynamics. 
Using a soluble {\it microscopic} model in string theory,
in which we can rigorously derive an effective hydrodynamic description, we 
established that the first-order magneto-hydrodynamics 
constructed from the positivity of the entropy current
is inconsistent. 
A careful analysis of conformal invariance of the effective 
hydrodynamic description allowed us to identify additional dissipative terms. 
We showed that the modified magneto-hydrodynamics 
provides a consistent interpretation of AdS/CFT results. 
On the other hand, we point out that there is no conflict between our 
results and the HKMS approach~\cite{Kovtun} who considered 
the case of the small magnetic field in the hydrodynamic limit. 

{\it Acknowledgements}.
The authors would like to thank Sam Vazquez for valuable discussions.
Research at Perimeter Institute is supported by the
Government of Canada through Industry Canada and by the Province of
Ontario through the Ministry of Research \& Innovation. AB
gratefully acknowledges further support by an NSERC Discovery grant
and support through the Early Researcher Award program by the
Province of Ontario.

%%%%%%%%%%%%%%%%%%%%%%%%%%%%%%%%%%%%%%%%%%%%%%%%%%%%%%%%%%%%%%%%%%%%%%%%%%%%%%%%%%%%%%%%%%%%%%%%%%

\end{document}